\documentclass[aps, prb, twocolumn,floatfix, showpacs, 10pt]{revtex4}
\usepackage{amsmath}
\usepackage{amsfonts}
\usepackage{amsbsy}
\usepackage{amssymb}
\usepackage{graphicx}
\usepackage{textcomp}
\usepackage[caption=false,singlelinecheck=false]{subfig}
\usepackage{xcolor}
\usepackage{mathrsfs}
\usepackage{bm}
\usepackage{ulem}
\usepackage[colorlinks,linkcolor=red,citecolor=blue,filecolor=magenta,
urlcolor=cyan]{hyperref}

\allowdisplaybreaks

\begin{document}
\title{Magnetic Chern Insulators in a monolayer of Transition Metal Trichalcogenides}

\author{Archana Mishra}

\author{SungBin Lee}

\affiliation{Korea Advanced Institute of Science and  Technology, Daejeon, South Korea}

\date{\today}
\begin{abstract}
A monolayer of transition metal trichalcogenides has received a lot of attention as potential two dimensional magnetic materials. The system has a honeycomb structure of transition metal ions, where both spin-orbit coupling and electron correlation effect play an important role.  
Here, motivated by these transition  metal series with effective doping or mixed valence case, we propose the possible realization of magnetic Chern insulators at quarter filled honeycomb lattice. We show that the interplay of intrinsic spin-orbit coupling and electron correlation opens a wide region of ferromagnetic Chern insulating phases in between metals and normal insulators. Within the mean field approximation, we present the phase diagram of a quarter filled Kane-Mele Hubbard model and also discuss the effects of Rashba spin-orbit coupling and nearest neighbor interactions on it. 
\end{abstract}
\pacs{71.10.Fd, 71.27.+a, 71.30.+h}
\maketitle


The effect of spin-orbit coupling plays an important role in the electronic structures of solids. In particular, spin-orbit coupling is essential for the realization of topological insulators where the gapless edge states are protected by time reversal symmetry \cite{kane2005,kane2005_1,fu2007,fu2007_1,buttiker2009,roy2009,moore2010,hasan2010,qi2010,ren2016}. Both theoretical prediction \cite{fu2007_1,bernevig2006, wu2008} and experimental  realization \cite{chen2009,konig2007,hsieh2008,hsieh2009,xia2009,hsieh2009_1} of topological insulators in real materials have been extended to the study of topologically non-trivial phases. More recently, it is also pointed out that the electron interaction effect could induce exotic phases such as topological Mott insulators.\cite{pesin2010,liu2016} Thus, the interplay of spin-orbit coupling and electron interaction has garnered a lot of attention, leading to a new discovery of materials and theoretical studies\cite{zheng2011,hohenadler2012,hung2013,laubach2014,griset2012,vaezi2012,hohenadler2014,
yu2011,meng2014,hohenadler2011,chung2014,budich2012,rachel2010,wu2012,lang2013,araujo2013,wen2011,gong2017discovery,kim2017charge}.

Topologically distinct phases introduced by Haldane showed that the quantum Hall phenomena could occur purely from the band structure in the absence of any external magnetic field, as a realization of the parity anomaly in (2+1) dimensional relativistic field theory \cite{haldane1988}. By introducing the staggered flux on a honeycomb lattice, the system becomes an insulator with a non-zero topological invariant, termed as a Chern insulator. Experimental realization of Haldane model has been recently proposed in ultra cold atom system \cite{jotzu2014}, yet none of them has been reported in any solid state systems.

Here, we propose possible realization of Chern insulators in two dimensional van der Waals materials, especially in transition metal trichalcogenides.
The van der Waals materials are characterized with layered crystals where  individual layers are weakly coupled via van der Waals forces but with strong covalent bonding in the layer. Thus it is possible to peel away a single layer breaking the van der Waals bonds. One of the most well known examples is a single layer of graphene, peeled away from bulk graphite using scotch tape
\cite{geim2007,geim2013,novoselov2016}. Using scotch tape technique, a variety of van der Waals materials have been successfully exfoliated into atomically thin layers. 
It turns out that pure two dimensional materials are not just limited to graphene, rather, 2D hexagonal boron nitride and the family of transition-metal chalcogenides are also present \cite{wang2012,xu2014,chittari2016,sivadas2015,du2015,lin2016,kuo2016,lee2016}. 
In particular, the transition-metal trichalcogenide (TMTC) series (with chemical formula TMBX$_3$ where TM represents transition metals, B=P,Si or Ge and X represents chalcogens) are recently receiving a great attention in both theoretical studies \cite{chittari2016,sivadas2015,sugita2017,sugita2017electronic} and experiments \cite{li2013,du2015,lin2016,kuo2016,lee2016,park2016opportunities}. 

In TMTCs, the transition metal ions form a layered honeycomb structure, thus, a single 2D unit consisting of these transition metal atoms has similar lattice structure as that of graphene. However, unlike the case of graphene which has a zero bandgap, TMTC series have a sizable variation of bandgap ranging from 0.5 eV to 3.5 eV depending on the transition metal atoms \cite{bullett1979}. In addition, the transition metal compounds possess large spin-orbit coupling and strong electron correlations compared to the case of graphene. Hence, these monolayers of TMTC series open a whole zoo of new exotic phases in two dimensional honeycomb lattice allowing possible control of both electron interaction and spin-orbit coupling. So far, there have been many recent studies on TMTC materials especially with $3d$ transition metal ions but not much attention on TMTCs with $4d$ and $5d$ transition metal ions.

In this paper, we study the interplay of spin-orbit coupling and electron correlation motivated by TMTC materials with $4d$ and $5d$ transition metal ions. Especially, we study quarter- (or three-quarter) filled system with effective pseudospin-1/2 model.  
As a minimal model, we consider the Kane-Mele Hubbard model \cite{zheng2011,hohenadler2012,hung2013,laubach2014,griset2012,vaezi2012,hohenadler2014,
yu2011,meng2014,hohenadler2011,chung2014,budich2012,rachel2010,wu2012,lang2013,araujo2013,wen2011}.
At quarter filling, we found several metallic and insulating phases within mean field approximation; ferromagnetic Chern metals, ferromagnetic Chern insulators and ferromagnetic normal insulators with broken time reversal and inversion symmetries. In particular, we point out that the magnetic Chern insulators could naturally arise when both spin-orbit coupling and electron interactions are present at quarter filling. In addition, we also found the possible transition from Chern insulator to normal insulator as originally proposed in the Haldane model\cite{haldane1988}. We also investigate the stability of these phases in the presence of nearest-neighbor interaction and Rashba spin-orbit coupling.

In $4d$ or $5d$ TMTCs, the presence of strong spin-orbit coupling and crystal field splitting can split $t_{2g}$ orbitals of transition metals ions (octahedral sites) into lower quartet orbitals with the effective total angular momentum $j=3/2$ and upper doublet with $j=1/2$ in the atomic limit \cite{sugano2012, enda2013}. When there are 9 or 11 electrons per unit cell (two sites) of honeycomb lattice, the $j=3/2$ orbitals are fully filled, while the $j=1/2$ orbitals of two sites have one or three electrons in total, resulting in effective quarter- or three-quarter fillings with pseudospin-1/2 model.  
Such fillings that include odd number of electrons per unit cell, can be realized by mixed valence of transition metal ions in two different sublattices, for instance $d^4$ (or $d^6$)  and $d^5$ in each sublattice. It can be also realized by doping via gating or hydrogen substitution.

Before we study the quarter filled case, we briefly summarize the earlier work related to the Kane-Mele (Hubbard) model. The Kane-Mele model was first proposed to study the quantum spin Hall (QSH) effect in graphene, but, due to very small spin-orbit coupling, the topological properties were not clearly visible. Instead, the search was extended
to real materials with strong spin-orbit coupling \cite{pesin2010,jackeli2009,shitade2009,wunderlich2005,moon2013,sinova2004}. There were also studies of possible QSH phases due to the spontaneous spin SU(2) symmetry breaking, induced by electron interactions even in the absence of spin-orbit coupling \cite{raghu2008,weeks2010}. Further related work on topological phase transitions in the presence or absence of spin-orbit coupling have been studied in other lattices like kagome, decorated honeycomb and diamond etc  \cite{guo2009,ruegg2010,wen2010,smith2012}.
The electron correlation effects on the Kane-Mele model were also extensively studied at half filling  \cite{zheng2011,hohenadler2012,hung2013,laubach2014,griset2012,vaezi2012,hohenadler2014,
yu2011,meng2014,hohenadler2011,chung2014,budich2012,rachel2010,wu2012,lang2013,araujo2013}.
Away from half filling, possible pairing mechanism of superconductivity has received attention which could occur at  3/8 or 5/8 filling near the Van-Hove singularity in doped Kane-Mele model \cite{wen2011,ma2015,fukaya2016}. However, few studies related to the interplay of strong intrinsic spin-orbit coupling and electron correlations have been discussed for the case of 1/4 or 3/4 filling \cite{PhysRevB.85.073103} and there are no detailed study  of the full  phase  diagram at  these fillings.

We start by introducing the Kane-Mele Hubbard model. The Hamiltonian is,
\begin{align} 
\label{eq:1} 
\mathcal{H}=&-t\sum_{\langle ij\rangle, \alpha}\left(c^\dagger_{i \alpha}c_{j \alpha}+h.c.\right) \\
&-i\lambda_{so}\sum_{\langle\langle ij\rangle\rangle, \sigma}
\left(c^\dagger_{i \alpha} \nu_{ij} \sigma^z_{\alpha \beta} c_{j \beta}+h.c\right)\nonumber
+U\sum_{i}n_{i\uparrow}n_{i\downarrow}, 
\end{align}
where $c_{i \alpha}^{\dag} $ ($c_{i\alpha}$) is the electron creation (annihilation) operator at site $i$ with spin $\alpha \in \{ \uparrow, \downarrow \}$ on a honeycomb lattice, $n_{i\alpha}=c_{i\alpha}^\dag c_{i \alpha}$ is the number density operator, $\sigma^z$ is a Pauli matrix, $\langle i j \rangle$ and $\langle \langle i j \rangle \rangle$ denotes pairs of nearest-neighbor and next-nearest-neighbor sites respectively. $t$, $U$ and $\lambda_{so}$ are the nearest-neighbor hopping energy, the strength of the on-site Coulomb repulsion and the second-neighbor spin-orbit coupling strength respectively. Throughout this paper, we set the hopping amplitude $t\equiv 1$.
$\nu_{ij}= -\nu_{ji}=\pm1$, depending on whether the electron traversing from $i$ to $j$ makes a right (+1) or a left (-1) turn.

At quarter filling, the system remains metallic with or without spin-orbit coupling for the non-interacting case $U=0$. At this filling, irrespective of the spin-orbit coupling strength, the perfect nesting wave vectors are absent at the Fermi surface. Thus, we neglect the instability of any charge density wave or spin density wave with finite momentum when the onsite repulsion $U$ is turned on. 
The interaction term in the Hamiltonian Eq.~\eqref{eq:1} can be rewritten in terms of spin operator $\bm{{S}}= c^\dagger_{i \alpha}\frac{\bm{\sigma}_{\alpha \beta}}{2}c_{i \beta}$ and the total number of electrons in the system is $N_e$; 
$U\sum_{i}n_{i\uparrow}n_{i\downarrow}=- \sum_{i} \frac{2U}{3}\bm{{S}}^2_i+\frac{U N_e}{2}$.
The last term $\frac{U N_e }{2}$ is a constant and just shifts the total energy of the system, thus can be ignored. We solve this interacting Hamiltonian using mean field approximation and the mean field Hamiltonian can be written as,
\begin{equation}
\label{eq:4}
 \mathcal{H}_{MF}=\mathcal{H}_0+\sum_{i}\bm{M}_i\cdot\bm{{S}}_i+\frac{3}{8U}\sum_i(\bm{M}_i)^2
\end{equation}
where $\mathcal{H}_0$ is the non-interacting part of the Hamiltonian and $  \bm{M}_i=-\frac{4U}{3}\langle\bm{{S}}_i\rangle$. 
There are two vector order parameters represented by $\bm{ M}_{a}$ where $a$ labels the two sublattices ($A,~B$) of honeycomb lattice. 
This mean-field Hamiltonian $\mathcal{H}_{MF}$ has the following form up to a constant term
\begin{eqnarray}
\label{eq:9}
\mathcal{H}_{MF} &=& \sum_{\bm k}c^\dagger_{\bm{k}}~ {h}_{MF} (\bm{k}) ~ c_{\bm{k}},  \\
h_{MF}(\bm{k}) &=& d_{1} (\bm{k}) ~ \tau^x \otimes \mathbb{I}+d_{2} ( \bm{k})~ \tau^y\otimes \mathbb{I} + d_{3}(\bm{k}) ~ \tau^z \otimes\sigma^z  \nonumber \\
 &+& \sum_{\mu \in \{ x,y,z \}} \Big( M^\mu \mathbb{I} \otimes \sigma^\mu + m^\mu \tau^z \otimes\sigma^\mu \Big)
\end{eqnarray}
where $c_{\bm{k}} = (c_{A\uparrow}(\bm{k}), c_{B\uparrow}(\bm{k}), c_{A\downarrow}(\bm{k}), c_{B\downarrow}(\bm{k}) )^T$ is the basis for honeycomb ($A,~ B$) sublattices with spins $\uparrow,\downarrow$. The Hamiltonian matrix $h_{MF} (\bm{k})$ can be represented in terms of two Pauli matrices $\bm{\tau} = (\tau^x, \tau^y, \tau^z)$ and 
$\bm{\sigma} = (\sigma^x, \sigma^y, \sigma^z)$ for $A,~B$ sublattices and spin $\uparrow,~\downarrow$ respectively.  
$ d_{1} (\bm{k}) = (1+ \cos k_1+ \cos k_2)$, 
$ d_{2} (\bm{k}) = (\sin k_1 - \sin k_2  )$ and 
$ d_{3} (\bm{k}) = 2 \lambda_{so}( \sin k_1 + \sin k_2 - \sin (k_1+ k_2))$ with $k_1, ~k_2$ being the momentum components along the basis vectors $\bm{\hat e}_1$ and $\bm{\hat e}_2$ in a honeycomb lattice and $M^\mu \equiv (M_{A}^\mu + M_{B}^\mu)/4$, $m^\mu \equiv (M_{A}^\mu - M_{B}^\mu)/4$ with $\mu = x,y,z$. 


To solve a self-consistent equation for ${ \bm{M}}$, 
(i) start the iteration with a
random initial guess for each components of $\bm{M}_{a}$, (ii) diagonalize $H_{MF}$ using 
$\bm{M}_{a}$, (iii) tune 
the chemical potential to quarter filling by fixing the number of particles, 
(iv) calculate the expectation value of the spin vector
on each site in the unit cell and compute the new values of  $\bm{M}_{a}$. The whole process from step (ii) to (iv) is repeated
until all the quantities converge.  We repeat this process for various initial
guesses and sometimes find different mean field solutions. Comparing the energies of these solutions, we pick up the lowest energy state as the ground state of the interacting Hamiltonian. 


\begin{figure}
\begin{center}
\includegraphics[width=\linewidth,keepaspectratio]{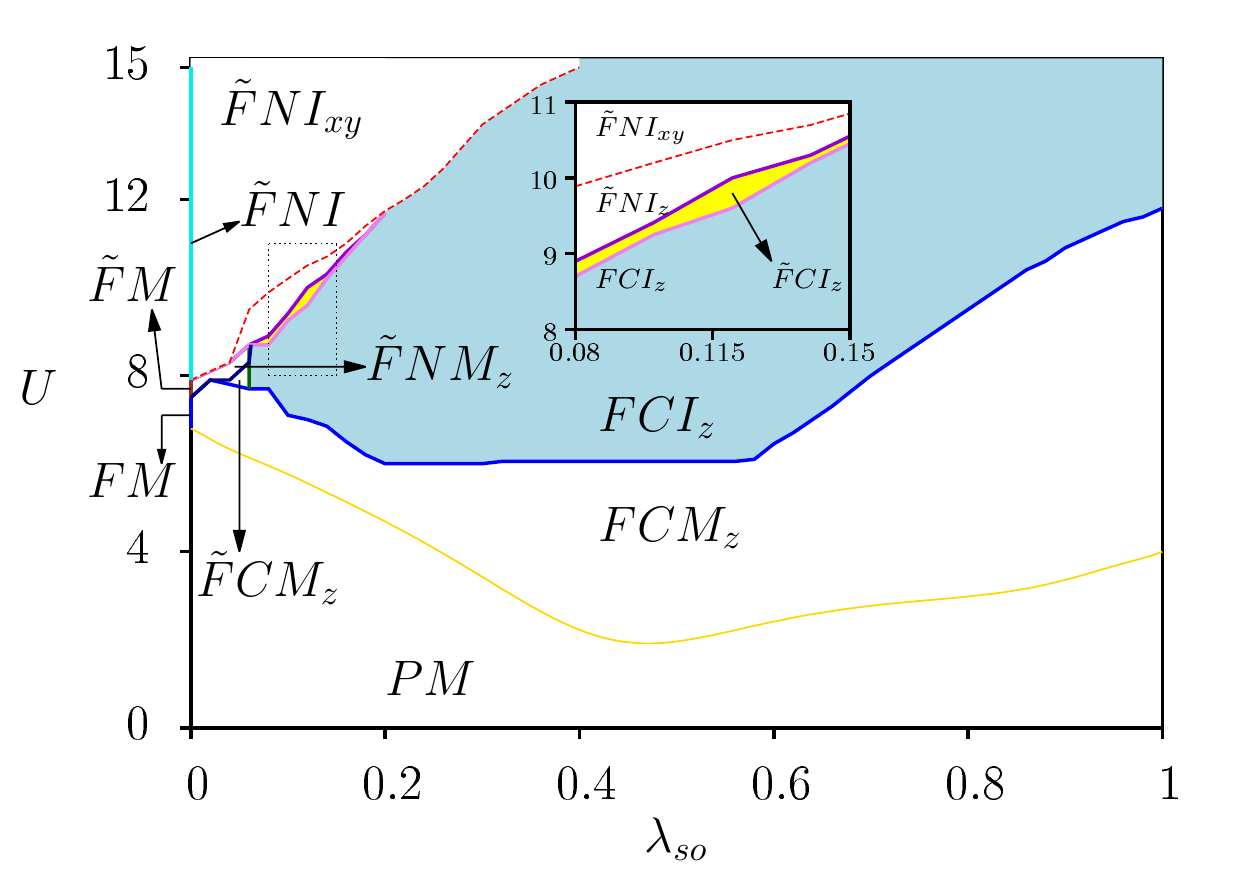}
\caption{\label{fig:1} (Color Online) Phase diagram for the Kane-Mele Hubbard model at quarter filling as functions of intrinsic spin-orbit coupling $\lambda_{so}$ and onsite Coulomb repulsion $U$. Here $t=1$.
The inset shows enlarged view of dashed box in the main plot. 
$PM$ : paramagnetic metal, $FM$ : ferro-magnetic (FM) metal, $FNI$ : FM normal insulator, $FCM$ : FM Chern metal, $FCI$ : FM Chern insulator. $F$ and $\tilde F$ distinguish FM order with or without inversion symmetry and the subscripts $xy$ or $z$ represent the direction of magnetic order. The blue and yellow shaded regions show the magnetic Chern insulator phases.
Detailed explanation of each phase is described in the main text.}
\end{center}
\end{figure}
Fig.~\ref{fig:1} is the phase diagram as a function of $U$ and $\lambda_{so}$ at quarter filling, based on solving the self consistency equation.
In the absence of both spin-orbit coupling and onsite interaction ($\lambda_{so} = U =0$), the system is in a metallic phase. With increasing $U$ but $\lambda_{so}=0$, the magnetic moment is being developed and the system goes into magnetically ordered phases. 
In the range $6.8<U<7.5$, the ferromagnetic metal is stabilized where $\bm{M}_A=\bm{M}_B\neq 0$ with broken time reversal symmetry. In this phase, $h_{MF} (\bm{k})$ is represented as two copies of graphene Hamiltonian with spin up and down, and  their energies are separated proportional to the magnetization values $\bm{M}_A=\bm{M}_B$. At quarter filling, hence, the lowest energy band remains gapless {\sl i.e.} metallic. At $U=U_c=7.5$, there is a second order phase transition into a 
ferromagnetic metal where  the system starts developing magnetization $\bm{M}_A \neq \bm{M}_B$ with broken inversion symmetry.
In this case, both $\bm{M} $ and $\bm{m}$ in Eq.~\ref{eq:9} are non-zero and thus, the lowest two energy  bands are separated at every momentum value, but, the bands still cross the Fermi level. On further increasing  $U$, the magnetization keep increasing opening a band gap  between the  lowest two bands  and a ferromagnetic insulator with inversion symmetry broken is stabilized. All of these phases at $\lambda_{so}=0$ are topologically trivial cases, thus we labeled these phases as paramagnetic metal `$PM$', ferromagnetic metal with inversion symmetry `$FM$',
ferromagnetic metal with inversion  symmetry broken `$\tilde FM$' and ferromagnetic normal insulator with broken inversion symmetry `$\tilde FNI$' as shown in Fig.~\ref{fig:1}.


In the presence of spin-orbit coupling ($\lambda_{so} \neq 0$), the non-interacting Kane-Mele model is just two copies of Haldane model with the phase factor $\phi=\pi/2$ discussed in Ref.~\onlinecite{haldane1988} and opposite sign for spin up and down. Here, there is no extra mass term related to inversion symmetry breaking in the Hamiltonian and the system is metallic at quarter filling. With increasing interaction $U$, the magnetization is being developed along $z$ direction {\sl i.e.} $M^z_A = M^z_B \neq 0$ above $U_c$ which depends on the value of $\lambda_{so}$. In Eq.~\eqref{eq:9}, this is equivalent to $M^z \neq 0$ and $M^{x,y}$ and $\bm{m}$ are zero. The preference of magnetization along $z$ direction can be easily understood by comparing the energies of single particle mean field Hamiltonian $h_{MF} (\bm{k})$ for two different cases, $M_A^z = M_B^z =M $ and $M_A^{x(y)} = M_B^{x(y)} =M$. 
In the former case, the single particle energy of the lowest band is $- \sqrt{d_1^2(\bm{k}) + d_2^2(\bm{k}) +d_3^2(\bm{k})} - M/2 $. 
In the latter case, the energy of the lowest band is $- \sqrt{ [(d_1^2(\bm{k})+ d_2^2(\bm{k}))^{\frac{1}{2}} +M/2]^2+d_3^2(\bm{k})}$.
Near the second order transition point from paramagnetic metal to ferromagnetic metal, the magnetization value $M$ is small and we see that the mean field solution with $M^z \neq 0$ has lower energy than the case with $M^{x(y)} \neq 0$, thus magnetic order along $z$ direction is favored. With a finite $M$, the degenerate bands are separated having non zero Chern number $\pm 1$. Although the lowest two bands are well separated at each momentum, the bands still cross the Fermi level at quarter filling, thus a ferromagnetic Chern metal phase, `$FCM_z$', is stabilized. 

\begin{figure}[h!]
\begin{center}
\subfloat[]{\label{fig:2a}\includegraphics[scale=0.075]{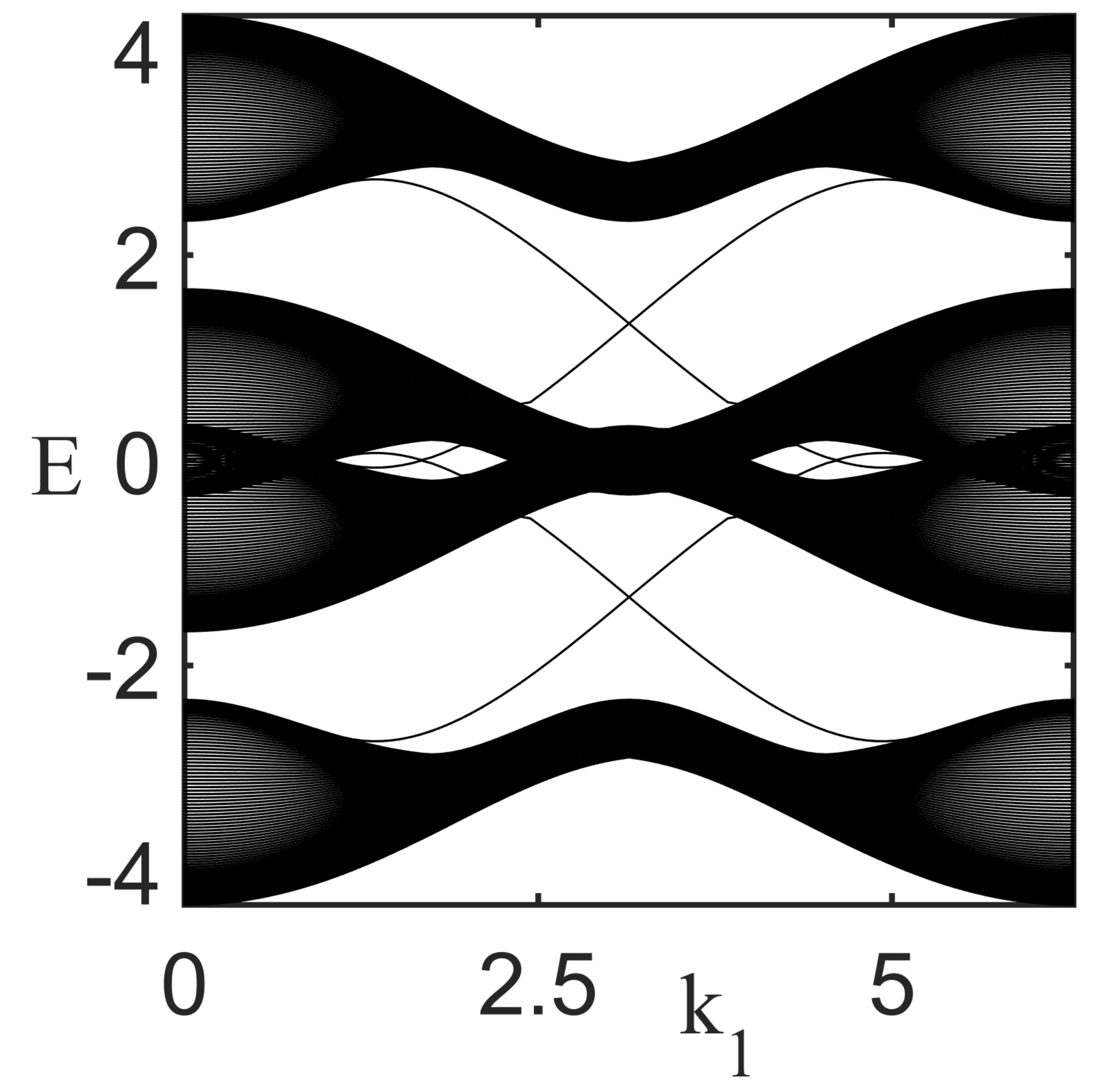}}~~
\subfloat[]{\label{fig:2b}\includegraphics[scale=0.075]{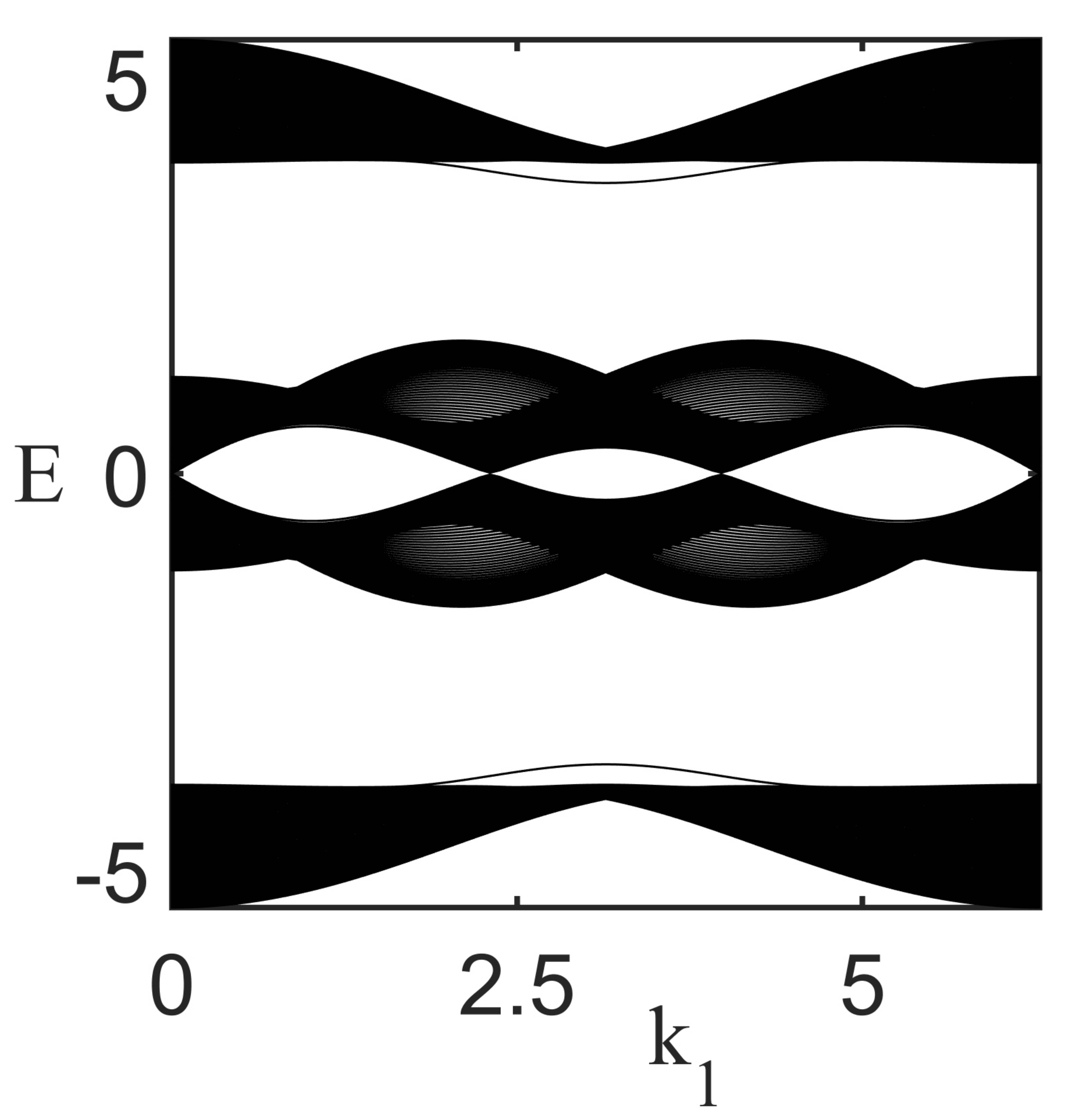}}
\caption{\label{fig:2} (Color Online) Energy spectra for system with zigzag edge for $\lambda_{so}=0.3$ and (a) $U=8$ which
is in $FCI_z$ region, (b) $U=13$ which is in $\tilde FNI_{xy}$ region.}
\end{center}
\end{figure}

With further increasing $U$, the lowest two bands are eventually separated, resulting in a second order phase transition from `$FCM_z$' phase to ferromagnetic Chern insulating phase denoted as `$FCI_z$' with $M^z \neq 0$ (See Fig.~\ref{fig:1}).
In this phase, quarter filling corresponds to filling the lowest band that has the Chern number $+1$, resulting in the Hall conductivity $\sigma_H = e^2 /h$. This Chern insulating phase can also be confirmed by the edge state calculation. 
We consider the honeycomb lattice with periodic boundary condition along one of the basis vector and zigzag edge along the direction of other basis vector. Based on the mean field Hamiltonian, we can plot the energy dispersion of the bands with edge states as shown in Fig.~\ref{fig:2}.
Fig.~\ref{fig:2a} shows the energy dispersion in $FCI_z$ at $U=8$ and $\lambda_{so}=0.3$. We see that there are gapless edge states for the $FCI_z$ phase along with the bulk energy gap thus indicating it to be a non-trivial  phase.
As shown in Fig.~\ref{fig:1}, it is remarkable that the presence of both spin-orbit coupling and Coulomb repulsion naturally opens a wide range of ferromagnetic Chern insulator at quarter filling. This is very distinct situation compared to the half filled case where the QSH phase exists even without electron correlation. 

For $0.06\leq\lambda_{so} \lesssim 0.2$, there is another second order phase transition within mean field approximation. Increase of $U$ leads to different magnetic phases where $M_A^z \neq M_B^z$, thus both $M^z$ and $m^z$ are non zero in Eq.\eqref{eq:9}. In this phase, the system still has non-trivial topology and becomes ferromagnetic Chern insulator with broken inversion symmetry denoted as `$\tilde FCI_z$'. With further increasing $U$, there exist a critical value of $m^z$ where the lowest two bands cross and then the system goes into a ferromagnetic normal insulator with broken inversion symmetry denoted as `$\tilde FNI_z$'. Such phase transitions between $FCI_z$ or $\tilde FCI_z$ and $\tilde FNI_z$ are exactly consistent with the phase transition in the original Haldane model that is induced by inversion breaking mass term at the phase $\phi = \pi/2$ \cite{haldane1988}. These phases and phase transitions are only stable for weak spin-orbit coupling. 
For $\lambda_{so}<0.06$, there are two metallic phases with broken inversion symmetry: ferromagnetic Chern metal 
`$\tilde{F}CM_z$' and ferromagnetic normal metal `$\tilde{F}NM_z$' as shown in Fig.~\ref{fig:1}.

For very large values of $U$, the system stabilizes a ferromagnetic normal insulator with magnetization in $xy$ plane but with broken inversion symmetry denoted as $\tilde FNI_{xy}$. (See Fig.\ref{fig:1}) 
The $\tilde FNI_{xy}$ phase can be understood by considering the spin-orbit coupling term as a perturbation in the large $U$ limit. In the absence of spin-orbit coupling, the single particle mean-field energies are exactly the same irrespective of whether the magnetization is along $z$ direction (case I) or $xy$ plane (case II) due to $SU(2)$ symmetry. When the spin-orbit coupling is small but finite, the second order correction in energy is $-4d_3^2 \Big( d_1^2 (\bm{k})+ d_2^2 (\bm{k}) \Big) \Big/ \Big(m^2 + 4 (d_1^2(\bm{k}) + d_2^2 (\bm{k})) \Big)^{3/2}$ for case I, whereas for case II, the correction is $-d_3^2 \Big/ \Big(M + \sqrt{m^2 + 4(d_1^2(\bm{k}) + d_2^2 (\bm{k}))} \Big)$. In the ferromagnetic phase with broken inversion symmetry, increase of $U$ leads to large $M \approx m$. Hence, the second order correction for the case I is proportional to $1/M^3$ and for the case II the correction is proportional to $1/M$. Therefore, at large $U$ limit, when the spin-orbit coupling term is present, the magnetization along $xy$ plane is preferred. 
For small $\lambda_{so} \lesssim0.2$, there is a first order phase transition from $\tilde FNI_z$ phase to $\tilde FNI_{xy}$, while for $\lambda_{so} \gtrsim 0.2$, first order phase transition is directly from $FCI_z$ phase to $\tilde FNI_{xy}$ phase. (See Fig.~\ref{fig:1}). Fig.~\ref{fig:2b} shows the energy band for the honeycomb lattice with zigzag edge in $\tilde  FNI_{xy}$  phase, at $U=13$ and $\lambda_{so}=0.3$. There is no edge state crossing at quarter filling which indicates the system in a trivial insulating phase.

Finally, we also discuss the stability of the phase diagram in the presence of Rashba spin-orbit coupling and the nearest neighbor Coulomb interaction. The nearest-neighbor Coulomb interaction favors charge order and develops a mass term which breaks the inversion symmetry of the lattice. Thus, the phase
with ferromagnetic order $\bm{M}^A\neq \bm{M}^B$ is further stabilized and the area of inversion broken ferromagnetic normal  insulator  phase is increased in the phase diagram.
When both intrinsic ($\lambda_{so}$) and Rashba ($\lambda_R$) spin-orbit couplings are present, the particle hole symmetry is broken. In this case, the energy dispersion for non-interacting Hamiltonian $\epsilon^{n}_{\bm k}(\lambda_{so},\lambda_R)$ is independent of the sign of  $\lambda_R$ but depends on the sign of $\lambda_{so}$; $n\in[1,4]$ is the band index from the lowest to highest energy bands.
Here, $\epsilon^{n}_{\bm k}(\lambda_{so},\lambda_R)=-\epsilon^{5-n}_{\bm k}(-\lambda_{so},\lambda_R)$. This can be easily seen from the energy form of the single particle Hamiltonian with both $\lambda_{so}$ and $\lambda_R$. 
We found that the  phase diagram  remains almost unchanged
for $\lambda_R\leq |\lambda_{so}|$ but the magnetization value decreases with the increase of $\lambda_R$. 
For $\lambda_R> |\lambda_{so}|$ and negative $\lambda_{so}$, the system in $\tilde FNI_{xy}$ phase undergoes a first order  phase transition to $FCI_z$ phase and the critical $\lambda_R$ value for this transition increases with increase in $U$. For $\lambda_R>\lambda_{so}$ with positive $\lambda_{so}$, there is a first order transition from $\tilde FNI_{xy}$ to $\tilde FNI_z$ phase. 
For the latter case, the system in $FCI_z$ phase goes to $\tilde FNI_z$ under first order phase transition for intermediate $U$. 


To summarize our results, we have studied the interplay of intrinsic spin-orbit coupling and onsite Coulomb interactions at quarter filled honeycomb lattice. Motivated by series of TMTCs with $4d$ and $5d$ transition metal ions, we have focused on the Kane-Mele Hubbard model at quarter filling and have shown possible realization of magnetic Chern insulators due to the presence of both electron interaction and spin-orbit coupling. Within mean field approximation, we found that a wide range of magnetic Chern insulators is naturally opened as shown in Fig.~\ref{fig:1}. Furthermore, it can also lead to the phase transition between Chern insulator and normal insulator that was originally proposed by Haldane  \cite{haldane1988}, by stabilizing magnetic order. We have also explored the stability of those phases in the presence of Rashba spin-orbit coupling and nearest neighbor Coulomb interaction. 
Fluctuation effect and the nature of phase transitions beyond mean field theory are also required as future studies. 
Another interesting aspect is to investigate the effect of magnetic field on these systems with spin-orbit coupling and considerable
electron correlation\cite{archana2017}. In the absence of Coulomb interaction, magnetic field can induce many interesting topological phases with high Chern numbers, studied in the Kane-Mele model \cite{beugeling2012}. The effect of interaction can lead to various exotic topological phases which is beyond the  scope of this paper. 


 The authors acknowledge supports from KAIST startup and National Research Foundation Grant (NRF-2017R1A2B4008097). A. Mishra is supported by BK21 plus. The authors would like to thank Prof. Leon Balents,  Prof.  S. R. Hassan, Dr. Dibyakrupa Sahoo and Dr. Vinu Lukose for their useful comments.
 

\end{document}